\documentclass[preprint,showpacs,preprintnumbers,amsmath,amssymb,apssamp,prb,aps]{revtex4}

\usepackage{graphicx}
\usepackage{dcolumn}
\usepackage{bm}

\begin{document}

\preprint{APS/123-QED}

\title{First-order spatial coherence of excitons in planar nanostructures:
a k-filtering effect}

\author{L. Mouchliadis}

\author{A.~L. Ivanov}
\affiliation{Department of Physics and Astronomy, Cardiff
University, Queens Buildings, CF24 3AA, Cardiff, UK}
\date{\today}

\begin{abstract}
We propose and analyze a $k_{\|}$-filtering effect which gives
rise to the drastic difference between the actual spatial
coherence length of quasi-two-dimensional (quasi-2D) excitons or
microcavity (MC) polaritons in planar nanostructures and that
inferred from far-field optical measurements. The effect
originates from the conservation of in-plane wavevector $k_{\|}$
in the optical decay of the particles in outgoing bulk photons.
The $k_{\|}$-filtering effect explains the large coherence lengths
recently observed for indirect excitons in coupled quantum wells
(QWs), but is less pronounced for MC polaritons at low
temperatures, $T \lesssim 10$\,K.
\end{abstract}



\maketitle

Long-range spatial coherence is a fingerprint of well-developed
Bose-Einstein (BE) statistics. Measurements of the first-order
spatial coherence function $g^{(1)}$ and the coherence length
$\xi$ have allowed to visualize the BE condensation transition in
a trapped Bose gas of Rb atoms \cite{Bloch}. There are several
recent reports on the observation of long-range spatial optical
coherence in a low-temperature quasi-2D system of microcavity
polaritons \cite{Kasprzak,Deng} and indirect excitons
\cite{Gorbunov,TimofeevB,Yang,Butov}. In this case, the resonant
optical decay of MC polaritons or QW excitons in bulk photon modes
allows to map the in-plane coherence function $g^{(1)}$ of the
particles, by measuring the optical coherence function ${\tilde
g}^{(1)}$ of the emitted photons. It is commonly assumed that the
coherence length of QW excitons (MC polaritons), $\xi_{\rm x}$
($\xi_{\rm p}$), associated with $g^{(1)}$, is identical to that,
$\xi_{\gamma}$, of the optical coherence function ${\tilde
g}^{(1)}$.

In this Letter, we report a $k_{\|}$--{\it filtering effect},
which can strongly influence the optical coherence function
${\tilde g}^{(1)}$ measured from a planar nanostructure, and
calculate $g^{(1)}$ and ${\tilde g}^{(1)}$ for QW excitons and MC
polaritons. For QW excitons, the $k_{\|}$-filtering effect
tremendously increases the optical coherence length
$\xi_{\gamma}$, leading to $\xi_{\gamma} \gg \xi_{\rm x}$, and can
naturally explain the $\mu$m coherence lengths observed for
indirect excitons and attributed to spontaneously developed
coherence. The effect is less pronounced for MC polaritons, still
with $\xi_{\gamma} \gtrsim \xi_{\rm p}$.

The $k_{\|}$-filtering effect stems from the energy and in-plane
momentum $\hbar k_{\|}$ conservation in the resonant conversion
``quasi-2D QW exciton (MC polariton) $\rightarrow$ outgoing bulk
photon''. For a (coupled) quantum well surrounded by thick
co-planar barrier layers, the case illustrated in Fig.\,1, only
low energy optically-active excitons from the radiative zone
$k_{\|} \leq k_0 = (\sqrt{\varepsilon_{\rm b}}/c)\omega_0$, with
$\varepsilon_{\rm b}$ the dielectric constant of (AlGaAs) barrier
layers and $\hbar \omega_0$ the exciton energy at $k_{\|}=0$, are
bright, i.e., can emit far-field light
\cite{Hanamura,Andreani,Citrin}. In a far-field optical experiment
with the detection angle $2 \alpha$ [see Fig.\,1\,(b)], the
fraction of QW excitons which contribute to the optical signal is
drastically reduced further to the wavevector band $\Delta k_{\|}$
given by $0 \leq k_{\|} \leq k_{\|}^{(\alpha)} =
(k_0/\sqrt{\varepsilon_{\rm b}}) \sin \alpha \ll k_0$. The
$\alpha$-dependent narrowing of the detected states results in an
effective broadening of the first-order spatial coherence function
${\tilde g}^{(1)}$. In addition, the sharp cutoff of the detected
states at $k_{\|} = k_{\|}^{(\alpha)}$ yields an unusual
oscillatory behavior of ${\tilde g}^{(1)}$. The $k_{\|}$-filtering
effect has no analogy in optics of bulk excitons or polaritons.

The first-order spatial coherence function $g^{(1)}$
[\onlinecite{GlauberB,ScullyB}] of quantum well excitons, at a
fixed time, is given by $g^{(1)}({\bf r}_{\|}',{\bf r}_{\|}'') =
G^{(1)}({\bf r}_{\|}',{\bf r}_{\|}'')/[G^{(1)}({\bf r}_{\|}', {\bf
r}_{\|}') G^{(1)}({\bf r}_{\|}'',{\bf r}_{\|}'')]^{1/2}$ with
$G^{(1)}({\bf r}_{\|}',{\bf r}_{\|}'') =
\langle{\hat{\Psi}}^\dag({\bf r}_{\|}') \hat{\Psi}({\bf
r}_{\|}'')\rangle$, where $\hat{\Psi}({\bf r}_{\|}') =
(1/\sqrt{S})\sum_{{\bf k}_{\|}} e^{i {\bf k}_{\|} {\bf r}_{\|}'}
B_{{\bf k}_{\|}}$, ${\bf r}_{\|}$ is the in-plane coordinate, $S$
is the area, and $B_{{\bf k}_{\|}}$ is the exciton operator. Thus
for isotropically distributed QW excitons one receives:
\begin{equation}
g^{(1)} = g^{(1)}(r_{\|}) = \frac{1}{2 \pi n_{\rm 2d}}
\int_0^{\infty} J_0(k_{\|}r_{\|}) n_{k_{\|}} k_{\|} d k_{\|} \, ,
\label{filt1}
\end{equation}
where $r_{\|} = |{\bf r}_{\|}'' - {\bf r}_{\|}'|$, $n_{\rm 2d}$ is
the concentration of particles, $n_{{\bf k}_{\|}} = \langle
B^{\dag}_{{\bf k}_{\|}} B_{{\bf k}_{\|}} \rangle$ is the
occupation number, and $J_0$ is the zeroth-order Bessel function
of the first kind. For a classical gas of QW excitons at thermal
equilibrium, Eq.\,(\ref{filt1}), with $n_{{\bf k}_{\|}}$ given by
the Maxwell-Boltzmann (MB) distribution function $n_{k_{\|}}^{\rm
MB}$, yields the well-known result:
\begin{equation}
g^{(1)} = g_{\rm cl}^{(1)}(r_{\|}) = e^{-\pi r_{\|}^{2}/
\lambda_{\rm dB}^{2}} \, , \label{gauss0}
\end{equation}
where the thermal de Broglie wavelength is given by $\lambda_{\rm
dB} = [(2 \pi \hbar^2)/(M_{\rm x} k_{\rm B} T)]^{1/2}$ with $T$
the temperature and $M_{\rm x}$ the exciton in-plane translational
mass. For helium temperatures, one estimates from
Eq.\,(\ref{gauss0}) the coherence length of MB-distributed
indirect excitons in coupled QWs as $\xi_{\rm x} \sim \lambda_{\rm
dB} \sim 0.1\,\mu$m.

\begin{figure} [t]
\includegraphics*[width=0.6 \textwidth]{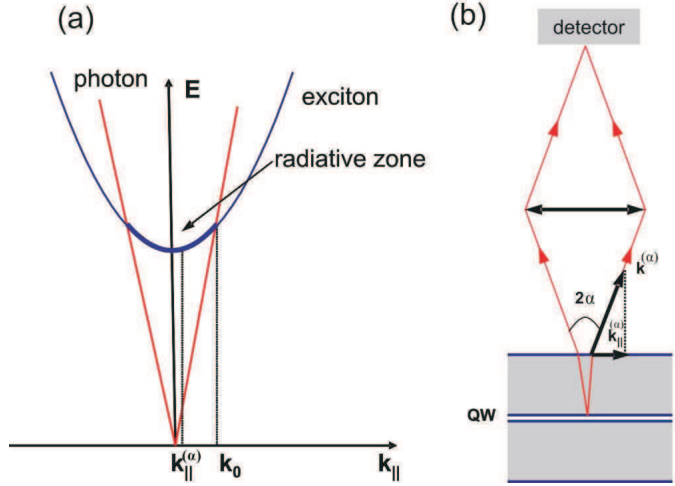}
\caption{(color online) Schematic of the $k_{\|}$-filtering
effect. (a) The exciton and photon dispersions. Only low-energy QW
excitons from the radiative zone $k_{\|} \leq k_0$ can emit
outgoing bulk photons. (b) A far-field optical experiment with the
detection angle $2\alpha$: A small fraction of QW excitons with
$|{\bf k}_{\|}| \leq k_{\|}^{(\alpha)} =
(k_0/\sqrt{\varepsilon_{\rm b}}) \sin \alpha$ contributes to the
optical signal.}
\end{figure}

Comparing with Eq.\,(\ref{filt1}), the spatial coherence function
${\tilde g}^{(1)}$ of photons emitted by QW excitons is given by
\begin{equation}
{\tilde g}^{(1)}(r_{\|} ) = \frac{\int_0^{\infty} G_{\rm
f}(k_{\|}) J_0(k_{\|}r_{\|}) n_{k_{\|}} k_{\|} d
k_{\|}}{\int_0^{\infty} G_{\rm f}(k_{\|}) n_{k_{\|}} k_{\|} d
k_{\|}} \, , \label{filter}
\end{equation}
where $G_{\rm f} = \Theta(k_{\|}^{(\alpha)} - k_{\|}) \Gamma_{{\rm
x}-\gamma}(k_{\|})$ is the $k_{\|}$-filtering function with
$\Theta(x)$ the step function and $\Gamma_{{\rm
x}-\gamma}(k_{\|})$ the efficiency of the resonant conversion of a
QW exciton in an outgoing bulk photon. The function $G_{\rm f}$
reduces the integration limits on the right-hand side (r.h.s.) of
Eq.\,(\ref{filter}) to the narrow band $\Delta k_{\|} =
[0,k_{\|}^{(\alpha)}]$ and describes the $k_{\|}$-filtering effect
in high-quality planar nanostructures. If both the function
$\Gamma_{{\rm x}-\gamma}(k_{\|})$ and the occupation number
$n_{k_{\|}}$ do not change significantly in the narrow band
$\Delta k_{\|}$, Eq.\,(\ref{filter}) yields:
\begin{equation}
{\tilde g}^{(1)} = {\tilde g}^{(1)}_{\rm f}(r_{\|}) = 2
J_1(k_{\|}^{(\alpha)}r_{\|})/(k_{\|}^{(\alpha)}r_{\|}) \, ,
\label{filter3}
\end{equation}
where $J_1$ is the first-order Bessel function of the first kind.
From Eq.\,(\ref{filter3}) one concludes that the optical coherence
length $\xi_{\gamma}$, evaluated as the half width at half maximum
of ${\tilde g}^{(1)} = {\tilde g}^{(1)}_{\rm f}(r_{\|})$, is given
by
\begin{equation}
4J_1(k_{\|}^{(\alpha)} \xi_{\gamma}) = k_{\|}^{(\alpha)}
\xi_{\gamma} \ \ \rightarrow \ \ k_{\|}^{(\alpha)} \xi_{\gamma}
\simeq 2.215 \, . \label{filter4}
\end{equation}
Equations (\ref{filter3})-(\ref{filter4}) illustrate the net
$k_{\|}$-filtering effect: $\xi_{\gamma}$ $\propto$
$1/k_{\|}^{(\alpha)}$ $\propto$ $1/\sin \alpha$ strongly increases
with decreasing aperture angle $2\alpha$. Below we analyze in more
detail the polarization function $g^{(1)}$ against the optical
${\tilde g}^{(1)}$, assuming no phase transition to a collective
(superfluid) quasi-2D state \cite{Popov}.

{\it First-order spatial coherence of non-interacting quasi-2D
bosons (excitons) in equilibrium.} In this case, the chemical
potential $\mu_{\rm 2d}$ is given by $\mu_{\rm 2d}^{(0)} = k_{\rm
B} T \ln (1-e^{-T_{0}/T})$ with $k_{\rm B}T_0 = (2
\pi/g)(\hbar^2/M_{\rm x})n_{\rm 2d}$ the quantum degeneracy
temperature and $g$ the spin degeneracy factor of bosons ($g=4$
for indirect excitons). By substituting $n_{\bf k_{\|}} = n^{\rm
BE}_{\rm k_{\|}}$ into Eq.\,(\ref{filt1}), where $n^{\rm BE}_{\rm
k_{\|}}$ is the Bose-Einstein occupation number, one receives:
\begin{eqnarray}
g^{(1)} &=& g^{(1)}_{\rm nint}(r_{\|}) = \frac{T}{T_{0}} \,
g_{1}\big(1 - e^{T_0/T}, \, e^{-\pi r_{\|}^{2}/\lambda_{\rm
dB}^{2}}\big)
\nonumber \\
&=& \frac{ T}{T_{0}} \sum_{n=1}^{\infty}\frac{ \big(1-e^{-T_{0}/T}
\big)^{n}}{n} \,\, e^{- \pi r_{\|}^{2}/ n \lambda_{\rm dB}^{2}} \,
. \label{corr2}
\end{eqnarray}
Here, the generalized Bose function \cite{Naraschewski}
$g_{\alpha}(x,y)$ with $\alpha = 1$ is defined as
$g_{\alpha}(x,y)= \sum_{k=1}^{\infty}(x^{k}y^{1/k})/k^{\alpha}$.

For small distances, $r_{\|} \ll \lambda_{\rm dB}$,
Eq.\,(\ref{corr2}) yields:
\begin{equation}
g^{(1)}(r_{\|} \ll \lambda_{\rm dB}) \simeq 1 - \frac{T}{T_{0}}
\frac{\pi r_{\|}^{2}}{\lambda_{\rm dB}^{2}}\,
\mbox{Li}_{2}(1-e^{-T_{0}/T}) \, , \label{corr3}
\end{equation}
where Li$_{\alpha}(x) = \sum_{k=1}^{\infty} x^k/k^{\alpha}$ with
$\alpha = 2$ is the polylogarithm. For $T \gg T_0$,
Eq.\,(\ref{corr3}) recovers the classical limit, $g^{(1)}_{\rm
cl}(r_{\|}\!\rightarrow\!0) \simeq 1 - (\pi
r_{\|}^{2})/\lambda_{\rm dB}^{2}$, which is consistent with
Eq.\,(\ref{gauss0}). For large distances,
$r_{\|}\!\gtrsim\!r_{\|}^{\rm (q)}\!=\!\lambda_{\rm dB}
\big[-(2/\pi)\ln(1 - e^{-T_0/T})\big]^{1/2}$, Eq.\,(\ref{corr2})
reduces to
\begin{equation}
g^{(1)}\big(r_{\|}\!\gtrsim\!r_{\|}^{\rm (q)}\big) \simeq
2\frac{T}{T_{0}}K_{0}\Bigg(\frac{r_{\|}}{r_0}\Bigg) \, ,
\label{corr4}
\end{equation}
where $K_{0}$ is the modified Bessel function of the second kind
and $r_{0}= \lambda_{\rm dB}/[-4\pi\ln(1-e^{-T_{0}/T})]^{1/2}$.
Equation~(\ref{corr4}) explicitly includes quantum corrections to
the first-order correlation function $g^{(1)}$, through $T_0
\propto \hbar^2$. For $r_{\|} \gg r_0$, Eq.\,(\ref{corr4}) reduces
further to the quantum limit:
\begin{equation}
g^{(1)} = g_{\rm q}^{(1)}(r_{\|}\,\gg\,r_0) = \sqrt{2 \pi} \,
\frac{T}{T_{0}} \sqrt{ \frac{r_{0}}{r_{\|}}} \, e^{-r_{\|}/r_{0}}
\, . \label{corr5}
\end{equation}
For temperatures $T \gg T_0$, the spatial coherence function is
well approximated by Eq.\,(\ref{gauss0}), and the quantum
corrections given by Eq.\,(\ref{corr5}) refer to large $r_{\|}
\gtrsim \lambda_{\rm dB} \sqrt{(2/\pi) \ln(T/T_0)} \gg
\lambda_{\rm dB}$, and, therefore, to very small values of
$g^{(1)}$. The latter conclusion is consistent with the $e^{-\pi
r_{\|}^2/n \lambda_{\rm dB}^2}$ -- series on the r.h.s. of
Eq.\,(\ref{corr2}). For $T \lesssim T_0$, when Bose-Einstein
statistics is well-developed, Eq.\,(\ref{corr5}) is valid for
distances larger than $\lambda_{\rm dB}
\sqrt{(2/\pi)}\,e^{-T_0/2T} \ll \lambda_{\rm dB}$, so that
$g^{(1)}$ is well-approximated by $g_{\rm q}^{(1)}$ for any
$r_{\|}$.

Thus, with temperature $T$ decreasing from $T \gg T_0$ to $T
\lesssim T_0$, the coherence function $g^{(1)}$ changes from the
$n_{\rm 2d}$-independent Gaussian $g_{\rm cl}^{(1)}(r_{\|})$,
given by Eq.\,(\ref{gauss0}), to the $n_{\rm 2d}$-dependent
exponentially decaying $g_{\rm q}^{(1)}(r_{\|})$, given by
Eq.\,(\ref{corr5}). The quantum statistical effects considerably
increase the correlation length $\xi_{\rm x}$, as shown in
Fig.\,2. For $T \lesssim T_0$ one has $\xi_{\rm x} \sim r_0 \simeq
[\lambda_{\rm dB}/(2 \sqrt{\pi})]e^{T_0/2T}$, i.e., $\xi_{\rm x}$
increases exponentially with increasing density $n_{\rm 2d}$. This
is due to large population of the low-energy states, in particular
the ground-state mode ${\bf k}_{\|}\!=\!0$: $n^{\rm BE}_{k_{\|}=0}
= e^{T_0/T} - 1$.

\begin{figure} [t]
\includegraphics*[width=0.5\textwidth]{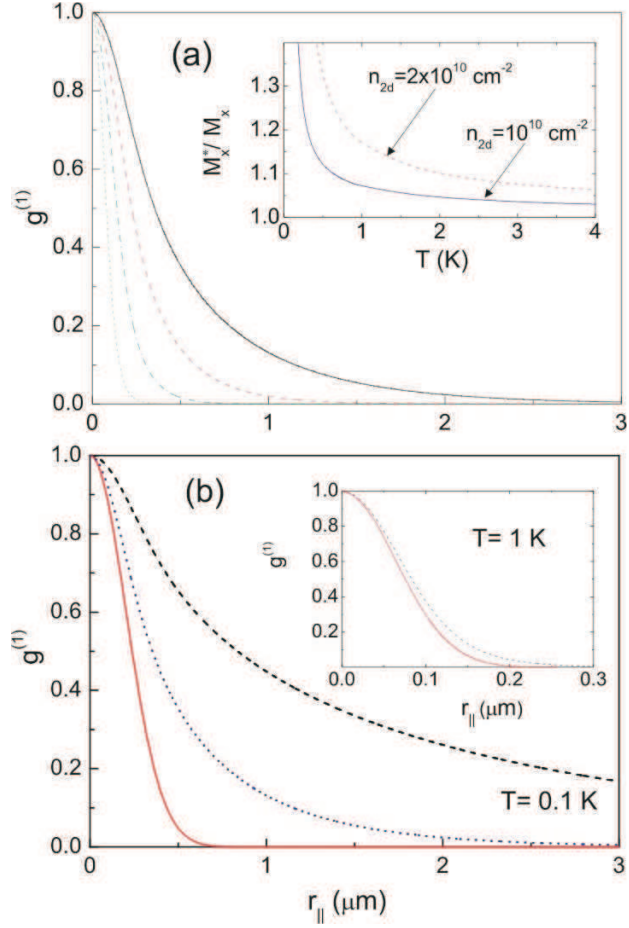}
\caption{(color online) (a) The first-order spatial coherence
function $g^{(1)} = g^{(1)}_{\rm ind}(r_{\|})$ of indirect
excitons in a GaAs coupled QW structure with $d_z = 11.5$\,nm and
$w = 15$\,nm: $n_{\rm 2d} = 10^{10}$\,cm$^{-2}$ and $T = 1$\,K
(dotted line), 0.4\,K (dash-dotted line), 0.2\,K (dashed line),
and 0.1\,K (solid line). Inset: The renormalized mass $M_{\rm
x}^*$ against temperature $T$, calculated with Eq.\,(\ref{mass})
for $n_{\rm 2d} = 10^{10}$\,cm$^{-2}$ (solid line) and $2 \times
10^{10}$\,cm$^{-2}$ (dashed line). (b) $g^{(1)} = g^{(1)}_{\rm
cl}(r_{\|})$ calculated with Eq.\,(\ref{gauss0}) (solid line),
$g^{(1)} = g^{(1)}_{\rm nint}(r_{\|})$ evaluated with
Eq.\,(\ref{corr2}), and $g^{(1)} = g^{(1)}_{\rm ind}(r_{\|})$
calculated with Eqs.\,(\ref{corr2}), (\ref{mu}) and (\ref{mass})
(dotted line): $n_{\rm 2d} = 10^{10}$\,cm$^{-2}$ and $T = 0.1$\,K.
Inset: the same functions evaluated for $n_{\rm 2d} =
10^{10}$\,cm$^{-2}$ and $T = 1$\,K.}
\end{figure}

{\it The coherence function $g^{(1)}$ of weakly-interacting
thermal QW excitons.} For circularly polarized excitons in a
single quantum well, the case relevant to MC polaritons, the
repulsive interaction between the particles is well approximated
by a contact potential $U_{\rm sqw} = (u_0/2) \delta({\bf
r}_{\|})$ with $u_0 = u_0^{\rm sqw} > 0$. In this case, the
mean-field (Hartree-Fock) interaction only shifts the chemical
potential, $\mu_{\rm 2d} = \mu_{\rm 2d}^{(0)} + u_0^{\rm sqw}
n_{\rm 2d}$, leaving unchanged Eqs.\,(\ref{corr2})-(\ref{corr5})
for the coherence function $g^{(1)}$.

For indirect excitons in coupled QWs, the mid-range dipole-dipole
repulsive interaction $U_{\rm cqw}$ of the particles cannot be
generally approximated by a contact potential. Following
[\onlinecite{Ulloa}], we use the two-parametric model potential
$U_{\rm cqw}(r_{\|}) = \big[ (\sqrt{\pi} u_0 w)/r_{\|}^3 \big]
\big(1 - e^{-r^2_{\|}/w^2}\big)$ with parameters $u_0 = u_0^{\rm
cqw} \simeq 4\pi (e^2/\varepsilon_{\rm b})d_z$
[\onlinecite{ButovR,Ivanov02}] and $w \simeq a_{\rm x}^{\rm
(2d)}$, where $\varepsilon_{\rm b}$ is the static dielectric
constant, $d_z$ is the distance between coupled quantum wells, and
$a_{\rm x}^{\rm (2d)}$ is the radius of an indirect exciton. The
model potential reproduces $1/r_{\|}^3$ behavior at $r_{\|}
\gtrsim a_{\rm x}^{\rm (2d)}$ and $1/r_{\|}$ Coulomb repulsive
potential at $r_{\|} \lesssim a_{\rm x}^{\rm (2d)}$. The
self-consistent Hartree-Fock (HF) analysis \cite{Fetter_Walecka}
of the Hamiltonian $H_{\rm x} = \sum_{{\bf p}_{\|}} [(\hbar^2
p_{\|}^2)/(2M_{\rm x})] B_{{\bf p}_{\|}}^{\dag} B_{{\bf p}_{\|}} +
1/(2S)\sum_{{\bf p}_{\|},{\bf l}_{\|},{\bf q}_{\|}} U_{\rm
cqw}({\bf q}_{\|}) B_{{\bf p}_{\|}}^{\dag}B_{{\bf l}_{\|}}^{\dag}
B_{{\bf l}_{\|}+{\bf q}_{\|}} B_{{\bf p}_{\|} - {\bf q}_{\|}}$
yields the $n_{\rm 2d}$- and $T$-dependent change of the in-plane
translational mass $M_{\rm x}$. In this case, $\mu_{\rm 2d}$ is
given by
\begin{eqnarray}
\mu_{\rm 2d} = \mu_{\rm 2d}^{(0)} &+& u_0 n_{\rm 2d} + \,
\frac{u_0}{2(\lambda^*_{\rm dB})^2} \, \Bigg[\frac{T_0^*}{T} +
\sqrt{\pi} \frac{w}{\lambda^*_{\rm dB}} \nonumber \\
&\times& \Big[ \frac{\sqrt{\pi}}{2} \frac{w}{\lambda^*_{\rm dB}}
\mbox{Li}_2 \big(1 - e^{-T^*_0/T}\big) - \mbox{Li}_{3/2} \big(1 -
e^{-T^*_0/T}\big) \Big] \Bigg] \, , \label{mu}
\end{eqnarray}
where, alongside Eq.\,(\ref{corr2}), both the de Broglie
wavelength $\lambda^*_{\rm dB}$ and the degeneracy temperature
$T_0^*$ are changed according to $M_{\rm x} \rightarrow M_{\rm
x}^*$. The particle mass $M_{\rm x}^*$ renormalized by the
dipole-dipole interaction is given as a single solution of the
transcendental equation:
\begin{equation}
\frac{1}{M_{\rm x}^{*}} = \frac{1}{M_{\rm x}} + \frac{u_0 w}{8
\sqrt{\pi} \hbar^2 \lambda^*_{\rm dB}} \left[\sqrt{\pi}
\frac{w}{\lambda^*_{\rm dB}} \frac{T_{0}^{*}}{T} - \mbox{Li}_{1/2}
\big(1 - e^{-T_{0}^{*}/T} \big)\right] \, . \label{mass}
\end{equation}

In Fig.\,2\,(a) we plot $g^{(1)} = g^{(1)}_{\rm ind}(r_{\|})$
evaluated numerically by using Eqs.\,(\ref{corr2}), (\ref{mu}) and
(\ref{mass}) for indirect excitons in a GaAs coupled QW structure.
In Fig.\,2\,(b), the coherence function $g^{(1)}_{\rm ind}$ is
compared with $g^{(1)}_{\rm cl}$ evaluated with
Eq.\,(\ref{gauss0}) and $g^{(1)}_{\rm nint}$ calculated with
Eq.\,(\ref{corr2}) for non-interacting excitons. The main result
is that the dipole-dipole repulsive interaction induces an
increase of the translational mass [see the inset of Fig.\,2\,(a),
note that the applied self-consistent HF approximation becomes
invalid when $\Delta M_{\rm x} = M_{\rm x}^* - M_{\rm x} \gtrsim
M_{\rm x}$] and, therefore, decreases the coherence length
$\xi_{\rm x}$ comparing to that of non-interacting particles [see
also Fig.\,3\,(a)]. The effect, however, becomes visible only at
temperatures well below 1\,K. For $T=1$\,K all three correlation
functions, $g^{(1)}_{\rm ind}$, $g^{(1)}_{\rm cl}$, and
$g^{(1)}_{\rm nint}$, nearly coincide, as is clearly seen in the
inset of Fig.\,2\,(b). In other words, for $n_{\rm 2d} \sim
10^{10}$\,cm$^{-2}$ and $T \gtrsim 1.5$\,K, which are relevant to
the experiments \cite{Gorbunov,TimofeevB,Yang,Butov}, the quantum
limit, i.e., $g^{(1)} = g^{(1)}_{\rm q}$ given by
Eq.\,(\ref{corr5}), cannot build up. For example, for $n_{\rm 2d}
= 10^{10}$\,cm$^{-2}$ and $T = 1.5$\,K one estimates $T_0 \simeq
T_0^* \simeq 0.65$\,K and $n_{k_{\|}=0}^{\rm BE} \simeq 0.54 < 1$,
so that BE statistics is rather weakly developed to influence the
coherence length $\xi_{\rm x}$.

\begin{figure} [t]
\includegraphics*[width=0.5\textwidth]{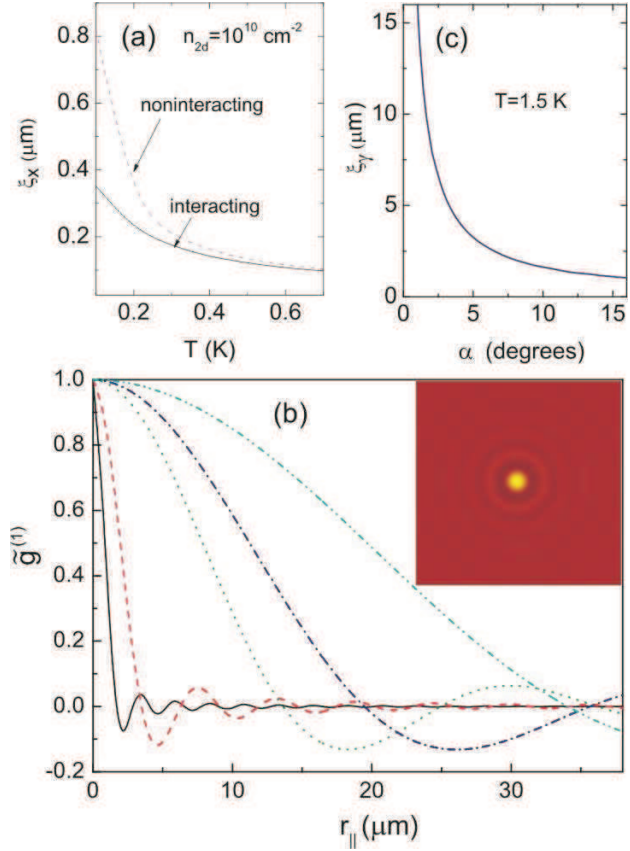}
\caption{(color online) (a) The dependence of the correlation
length $\xi_{\rm x}$ against temperature $T$, calculated for
noninteracting (dashed line) and dipole-dipole interacting (solid
line) indirect excitons. (b) The $k_{\|}$-filtering effect:
${\tilde g}^{(1)} = {\tilde g}^{(1)}(r_{\|})$ evaluated for
$\alpha = 18.9^{\circ}$ (solid line), $8.3^{\circ}$  (dashed
line), $2.1^{\circ}$ (dotted line), $1.4^{\circ}$ (dashed-dotted
line), and $0.8^{\circ}$ (dashed-double-dotted line). Inset: The
real-space 2D image of ${\tilde g}^{(1)}$. (c) The coherence
length $\xi_{\gamma}$ against the aperture angle $2\alpha$.}
\end{figure}

{\it The optical spatial coherence function ${\tilde g}^{(1)}$ of
indirect excitons.} In order to explain the experiments
\cite{Gorbunov,TimofeevB,Yang,Butov}, which demonstrate a
coherence length $\xi_{\gamma}$ much larger than $\xi_{\rm x} \sim
0.1\,\mu$m, we implement the concept of $k_{\|}$-filtering. In
this case, ${\tilde g}^{(1)} = {\tilde g}^{(1)}_{\rm ind}(r_{\|})$
is given by Eq.\,(\ref{filter}) with the efficiency of the
``indirect exciton $\rightarrow$ bulk photon'' conversion
$\Gamma_{{\rm x}-\gamma} = (2 k_0^2 - k_{\|}^2)/\big[k_0 (k_0^2 -
k_{\|}^2)^{1/2} \big]$
[\onlinecite{Hanamura,Andreani,Citrin,Ivanov97}]. In Fig.\,3\,(b),
we plot ${\tilde g}^{(1)}_{\rm ind}$ calculated for various
aperture angles, $2^{\circ} \lesssim 2\alpha \lesssim 40^{\circ}$.
The dependence ${\tilde g}^{(1)} = {\tilde g}^{(1)}_{\rm
ind}(r_{\|})$ is well-approximated by Eq.\,(\ref{filter3}). The
above approximation of ${\tilde g}^{(1)}$ by the ``device
function'' ${\tilde g}^{(1)}_{\rm f}$ is valid when $n_{k_{\|}} =
n^{\rm BE}_{E=\hbar^2k_{\|}^2/2M_{\rm x}}$ is nearly constant in
the rather narrow energy interval $0 \leq E \leq E^{(\alpha)}$,
i.e., when
\begin{equation}
E^{(\alpha)} = (\hbar k_{\|}^{(\alpha)})^2/2M_{\rm x} \ll k_{\rm
B} T e^{-T_0/T} \ . \label{filter5}
\end{equation}
For indirect excitons, the inequality (\ref{filter5}) with $T_0$
replaced by $T_0^*$ is definitely held for $n_{\rm 2d} \sim
10^{10}$\,cm$^{-2}$ and $T \sim 1$\,K (e.g., for $\alpha =
20^{\circ}$ the cutoff energy $E^{(\alpha)}$ is only
1.2\,$\mu$eV). Thus the $k_{\|}$-filtering effect yields the
correlation length $\xi_{\gamma} \simeq 2.215
\sqrt{\varepsilon_{\rm b}}/(k_0 \sin \alpha)$ with $k_0 \simeq 2.8
\times 10^5$cm$^{-1}$, according to Eq.\,(\ref{filter4}). As a
result, $\xi_{\gamma}$ is intrinsically scaled by the photon
wavelength, i.e., is in the $\mu$m length scale [see Fig.\,3\,(c),
where $\xi_{\gamma}$ is plotted against the angle $\alpha$].

Comparing to standard interference patterns in Young's double-slit
experiment, with contrast determined by ${\tilde g}^{(1)}$, the
oscillatory behavior of the optical coherence function ${\tilde
g}^{(1)} = {\tilde g}^{(1)}(r_{\|})$ is rather unusual [see
Eq.\,(\ref{filter3}) and Fig.\,3\,(b)]. This is a signature of the
$k_{\|}$-filtering effect: The $k_{\|}$-filtering function $G_{\rm
f} \propto \Theta (k_{\|}^{(\alpha)} - k_{\|})$ gives a sharp
cutoff at $k_{\|} = k_{\|}^{(\alpha)}$ in the integrals of
Eq.\,(\ref{filter}) that results in oscillations of ${\tilde
g}^{(1)}(r_{\|})$. In some aspects, the effect is similar to
Friedel oscillations in a Fermi liquid, with $\hbar
k_{\|}^{(\alpha)}$ akin to the Fermi momentum.

\begin{figure} [t!]
\includegraphics*[width=0.6\textwidth]{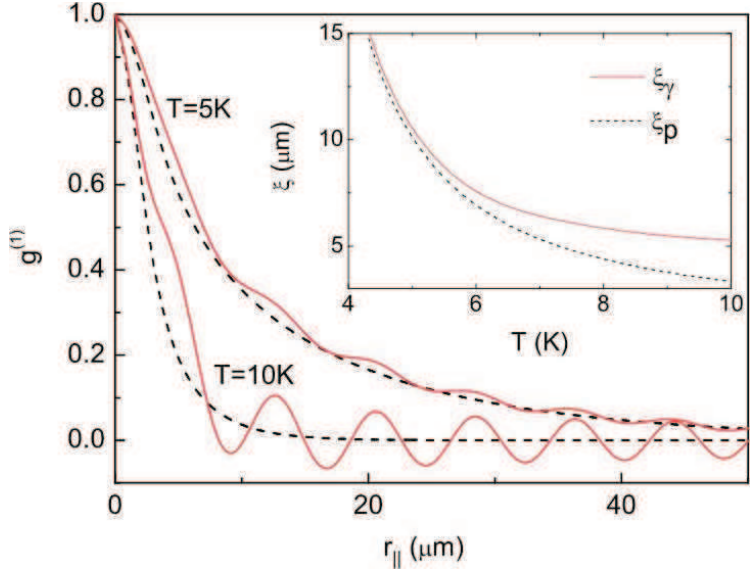}
\caption{(color online) The MC polariton coherence function
$g^{(1)} = g^{(1)}_{\rm MC}(r_{\|})$ (dashed lines) against that
of emitted photons, ${\tilde g}^{(1)} = {\tilde g}^{(1)}_{\rm
MC}(r_{\|})$ (solid lines). Inset: The coherence lengths $\xi_{\rm
p}$ ans $\xi_{\gamma}$ versus temperature $T$. The calculations,
which model the experiments \cite{Deng}, refer to a GaAs
microcavity with positive detuning $\delta = 7$\,meV and Rabi
splitting $\Omega_{\rm MC} = 4$\,meV. The density of MC polaritons
$n_{\rm 2d} = 10^8$\,cm$^{-2}$ and the aperture half-angle $\alpha
= 16.7^{\circ}$, so that $T_0 = 27.6$\,K and $E^{(\alpha)} =
0.96$\,meV. }
\end{figure}

{\it The coherence function ${\tilde g}^{(1)}$ of MC polaritons.}
In this case, the ``MC polariton $\rightarrow$ bulk photon''
conversion function in Eq.\,(\ref{filter}) is $\Gamma_{{\rm
x}-\gamma} = \Psi(k_{\|}) / \tau_{\gamma}(k_{\|})$ with $\Psi$ ($0
\leq \Psi \leq 1$) the photon component along a MC polariton
branch and $\tau_{\gamma}$ the radiative (escape) lifetime of a MC
photon. In Fig.\,4, $g^{(1)} = g^{(1)}_{\rm MC}(r_{\|})$
calculated with Eq.\,(\ref{corr2}) for circularly polarized MC
polaritons is compared with ${\tilde g}^{(1)} = {\tilde
g}^{(1)}_{\rm MC}(r_{\|})$ evaluated with Eq.\,(\ref{filter}).
According to the experiments \cite{Deng,Kasprzak}, we assume the
BE distribution of MC polaritons along the lower polariton branch
which is taken in the parabolic approximation with an effective
in-plane mass $M_{\rm MC}^{\rm lb}$. Comparing to the case of QW
excitons, the difference between $g^{(1)}_{\rm MC}$ and ${\tilde
g}^{(1)}_{\rm MC}$ is much smaller, still giving $\xi_{\gamma} >
\xi_{\rm p}$. This is because the cutoff energy $E^{(\alpha)}$ in
the $k_{\|}$-filtering effect is much larger than that relevant to
QW excitons [in Eq.\,(\ref{filter5}) $M_{\rm x}$ should be
replaced by $M_{\rm MC}^{\rm lb} \ll M_{\rm x}$]. The functions
$g^{(1)}_{\rm MC}$ and ${\tilde g}^{(1)}_{\rm MC}$ nearly
coincide, if $k_{\rm B}T \ll E^{(\alpha)}$ (see Fig.\,4).

We qualitatively explain a sharp increase of the coherence length
with decreasing temperature, found in the experiments with GaAs
coupled quantum wells \cite{Yang,Butov}, by combining the
$k_{\|}$-filtering effect with screening of disorder by
dipole-dipole interacting indirect excitons. The screening of the
random in-plane potential $U_{\rm rand}({\bf r}_{\|})$ can be
quantified by replacing $U_{\rm rand}$ with $U_{\rm eff} = U_{\rm
rand}({\bf r}_{\|})/[1 + (2/\pi) (u_0 M_{\rm
x}^*/\hbar^2)(e^{T_0^*/T} - 1)] \simeq U_{\rm rand}({\bf r}_{\|})
[(k_{\rm B}T)/(k_{\rm B} T + u_0 n_{\rm 2d})]$
[\onlinecite{Ivanov02}]: In high-quality GaAs coupled QWs the
screening process effectively develops at $T \lesssim 5$\,K,
giving rise to a well-defined single-particle momentum $\hbar {\bf
k}_{\|}$, as has been observed, e.g., in the experiments
\cite{Butov00,Parlangeli00,Butov01}. Thus the large correlation
length $\xi = \xi_{\gamma} \sim 1\,\mu$m of indirect excitons,
which strongly depends on $\alpha$, can naturally be explained by
the $k_{\|}$-filtering effect and can occur even for the
Maxwell-Boltzmann distributed particles. In order to see an
increase of $\xi$ due to quantum statistics, the bath temperature
should be decreased to tens of mK.

We appreciate valuable discussions with L.~V. Butov.

\end{document}